\documentclass[12pt]{iopart}

\usepackage{epsfig}

\begin{document}

\title{Global fluctuations in magnetohydrodynamic dynamos}

\author{V. Tanriverdi, A. Tilgner}

\address{Institute of Geophysics, University of G\"ottingen,
Friedrich-Hund-Platz 1, 37077 G\"ottingen}
\ead{vedat.tanriverdi@physik.uni-goettingen.de}
\ead{andreas.tilgner@physik.uni-goettingen.de}

\begin{abstract}
The spectrum of temporal fluctuations of total magnetic energy for several
dynamo models is different from white noise at frequencies smaller than the
inverse of the turnover time of the underlying turbulent velocity field.
Examples for this phenomenon are known from previous work and we add in this
paper simulations of the G.O. Roberts dynamo and of convectively driven dynamos in
rotating spherical shells. The appearance of colored noise in the magnetic
energy is explained by simple phenomenological models. The Kolmogorov theory of
turbulence is used to predict the spectrum of kinetic and magnetic energy
fluctuations in the inertial range.
\end{abstract}

\maketitle

\section{Introduction}

Several experiments have been carried out in the last decade in liquid sodium at
high magnetic Reynolds numbers and in highly turbulent flows. Measurements of
magnetic field fluctuations, either due to an externally imposed magnetic field
or due to magnetic field generated through a dynamo effect by the sodium flow
itself, generally reveal spectra with a $1/f$-noise at
low frequency \cite{Muller04, Gailit04, Bourgo02}.
Numerical simulations of magnetohydrodynamic (MHD) flows showed the same
phenomenon \cite{Ponty04, Dmitru07} although sometimes the precise form of the
low frequency noise varies \cite{Olson07}. The precise value of the exponent in
the
algebraic decay in a spectrum is not so important, but a behavior other than
white noise, i.e. colored noise, always invites closer
inspection \cite{Press78}. For example, a $1/f$-noise, frequently called flicker
 noise,
cannot extend down to zero
frequency if the power integral is to stay finite. The low frequency
cutoff to the $1/f$-spectrum should carry some informations about the dynamics
of an MHD dynamo.

Measurements of velocity in a fixed point of a turbulent flow frequently find
white noise at low frequency (see for example the data compilation in fig. 6.14
of \cite{Pope00}), but there are also examples of colored noise \cite{Ravele08}.
The simultaneous presence of colored noise in both magnetic and velocity
spectra looks like a plausible combination: If
flow velocities increase in a dynamo, the magnetic field is amplified more
rapidly and the amplitude of the magnetic field increases. If the velocity has
no white noise, so should the magnetic field have a nontrivial spectrum. In the
Karlsruhe experiment on the other hand \cite{Muller04} the $1/f$-fluctuations in
magnetic signals had no correspondence in pump pressures or volumetric flow
rates. This prompts us to investigate in more detail the mechanisms leading to
colored noise in magnetic spectra.

This paper will deal with the fluctuations of total magnetic energy, which is a
global measure of the field amplitude. A theory for the fluctuations of a global
quantity is of interest even though laboratory experiments usually report local
measurements. Numerical simulations always compute total
magnetic and kinetic energies. Their statistics are less noisy than those
of field amplitudes at a given point because of the spatial averaging inherent
to the computation of total energies. Spectra of the geomagnetic field are
frequently plotted as variations of another global quantity, the dipole moment
\cite{Consta05}, because the geomagnetic field is dominated by the dipole
component and low frequency variations reflect variations of the dipole moment.
Secular variations of the Earth's magnetic field again exhibit colored noise
\cite{Courti88, Consta07}. In astrophysics, we do have local measurements of
solar magnetic fields, but we only know a global amplitude for most stellar
magnetic fields.

The goal of this paper is to explore conditions leading to colored noise in the
total magnetic energy and to show that this type of noise can appear even if the
kinetic energy
has white noise. Three simplified models of MHD flows are presented in section \ref{models}.
The first two model the slow fluctuations per se, the last one by contrast
computes the spectrum of inertial range fluctuations in homogeneous and
isotropic turbulence. Section \ref{Roberts} presents numerical simulations of
the G.O. Roberts dynamo \cite{Robert72} and compares the results with section \ref{models}, whereas section
\ref{geo} does the same for simulations of convection driven dynamos in rotating
spherical shells.

\section{Phenomenological models \label{models}}

The basic problem in this paper is to find the temporal spectrum of a quantity
derived from a magnetic field $\bi B(\bi r,t)$, such as the magnetic energy. The
induction equation governs the evolution of $\bi B$ in a liquid conductor with
magnetic diffusivity $\lambda$ whose movement is given by the velocity field
$\bi v(\bi r,t)$:
\begin{equation}
\partial_t \bi B + \nabla \times (\bi B \times \bi v)
= \lambda \nabla^2 \bi B
\label{induc}
\end{equation}
Exact solutions of this equation are difficult to obtain, so that we resort
to phenomenological models.

\subsection{A single magnetic mode \label{single}}
Mean field magnetohydrodynamics have proven a most fruitful simplification of the
induction equation \cite{Krause80}. In this approach, the effect of small scale
fluctuations on the large scales are not computed exactly but are
modeled, in the simplest case as an $\alpha-$effect. The number of
magnetic degrees of freedom which need to be retained is thus reduced and in an
extreme simplification, only one mode remains. If we call $B$ the amplitude of
that mode, ${\tilde \alpha}(t)$ and $\beta$ the coefficients describing the
$\alpha-$effect and its quenching, respectively, and $\mu$ a coefficient related
to the magnetic dissipation, the simplest model reproducing the main features
of the induction equation is:
\begin{equation}
\partial_t B = {\tilde \alpha}(t) B - \beta B^3 - \mu B.
\label{alpha_model}
\end{equation}
${\tilde \alpha}$ is allowed to be time dependent in order to reflect a time
dependent velocity field. This time dependence will be essentially random for a
turbulent velocity field. The reduction of the $\alpha-$effect by the term
$\beta B^3$ models the retroaction of the magnetic field on the velocity field
via the Lorentz force (which is quadratic in the magnetic field)
in the Navier-Stokes equation. We now consider
$\alpha(t)={\tilde \alpha}(t)-\mu$ to be a random process with mean square 
$\langle\alpha^2\rangle$ and remove the dimensions from eq. (\ref{alpha_model}) by expressing
time in multiples of $\langle\alpha^2\rangle^{-1}$ and the magnetic field amplitude in
multiples of $(\langle\alpha^2\rangle/\beta)^{1/2}$. The adimensional quantities $t'$,
$\alpha'$ and $B'$ are given by $t'=t\langle\alpha^2\rangle$, $\alpha'=\alpha/\langle\alpha^2\rangle$
and $B'=B \sqrt{\beta/\langle\alpha^2\rangle}$. In the remainder of this section, all
quantities are understood to be nondimensional and the primes are omitted for
convenience. The nondimensional variables then  obey the equation:
\begin{equation}
\partial_t B = \alpha(t) B - B^3
\label{alpha_nondim}
\end{equation}
in which $\alpha(t)$ is a random variable with $\langle\alpha^2\rangle=1$. 
As long as $B$ is small, the solution to this equation is
\begin{equation}
B(t)=B(0) \exp \left( \int_0^t \alpha(\tau) d\tau \right).
\label{alpha_solution}
\end{equation}
For times small enough so that the exponent can be considered small, one has
\begin{equation}
\frac{B(t)-B(0)}{B(0)} \approx \int_0^t \alpha(\tau) d\tau.
\label{alpha_approx}
\end{equation}
Taking the Fourier transform of this equation, it follows that the spectrum of
$B$ is, apart from frequency independent prefactors, the same as the spectrum of
$\alpha$ divided by the square of the angular frequency, $\omega^2$. For
example, if the spectrum of $\alpha$ is a white noise, the spectrum of $B$
behaves as $\omega^{-2}$. For large times $t$,
eqs. (\ref{alpha_solution}) and (\ref{alpha_nondim})
become a poor approximation, which means that the $\omega^{-2}$ will not be
observable below some cutoff-frequency. If the mean of $\alpha$, $\langle\alpha\rangle$, is
different from zero, $B$ will be large enough for the nonlinear term
in eq. (\ref{alpha_nondim}) to become dominant after a time on the order of
$\langle\alpha\rangle^{-1}$. In that regime, and
concentrating on slow fluctuations, eq. (\ref{alpha_nondim}) reduces to 
$B^2 = \alpha$. Considering again the example of $\alpha(t)$ with a white noise,
one finds a spectrum of $B$ which is a white noise, too. The transition in the
spectrum of $B$ from $\omega^0$ to $\omega^{-2}$ occurs at a frequency which
increases with increasing $\langle\alpha\rangle$, because eq. (\ref{alpha_approx}) fails at
earlier times $t$.

\begin{figure}[h]
\begin{center}
\includegraphics[width=9cm]{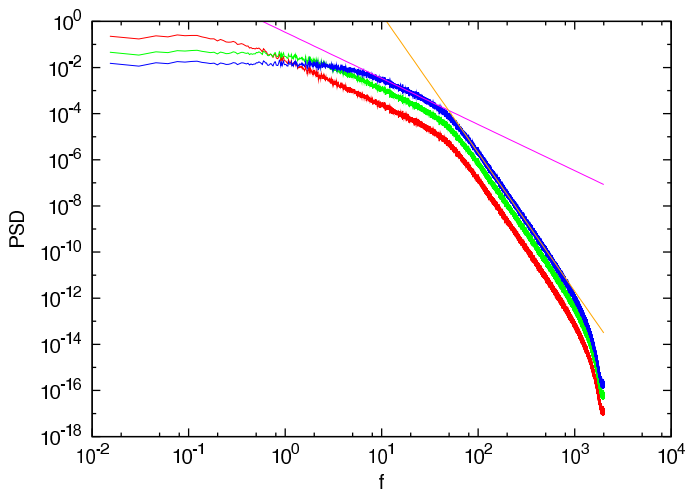}
\end{center}
\caption{Spectral power density of $B$, the solution of eq. (\ref{alpha_nondim}),
as a function of frequency $f=\omega/(2\pi)$ for
$<\alpha>=1$ (red), 5 (green) and 15 (blue). The straight lines indicate the power
laws $\omega^{-2}$ and $\omega^{-6}$.}
\label{fig_alpha}
\end{figure}

In order to test these ideas, we solved eq. (\ref{alpha_nondim}) numerically.
The random $\alpha(t)$ was generated by sending the output of a gaussian deviate
random number generator through a Butterworth filter \cite{Butter30, Rabine75}.
The filter was adjusted
such that its output had a spectrum as a function of frequency $f=\omega/(2\pi)$
in $1/(1+(f/f_1)^4)$ with $f_1=50$. The spectrum of $B$ is shown in
fig. \ref{fig_alpha} for different $\langle\alpha\rangle$. As expected,
this spectrum decays as $\omega^{-6}$ for $\omega > \omega_1 = 2\pi f_1$, and as
$\omega^{-2}$ for $\omega_c < \omega < \omega_1$. Below the cut-off $\omega_c$,
the spectrum is independent of $\omega$, and $\omega_c \propto \langle\alpha\rangle$.

Even though the spectra in fig. \ref{fig_alpha} differ only in the value of
$\omega_c$, the qualitative appearence of the underlying time series is quite
variable: The time series consists of intermittent bursts for small
$\langle\alpha\rangle$ and of random fluctuations around a well defined mean for
large $\langle\alpha\rangle$. A study of the probability distribution function
of the solutions of eq. (\ref{alpha_model}) together with a few graphs of
representative time series is presented in \cite{Leprov05}.

\subsection{Several magnetic modes \label{several}}
We will now investigate under which conditions the single mode model is
applicable to more general systems and extend the discussion of the previous
section to include several modes. It will be shown that the predictions of the
single mode model are recovered in the limit of small fluctuations.

The precise form of the dynamical system used as a model does not matter
much for the following analysis, but a specific system has to be chosen for the
numerical examples. In order to stay as close as possible to the previous
section, let us assume $\nabla \cdot \bi v=0$ and rewrite the left hand side of
eq. (\ref{induc}) as $\partial_t B_i + \sum_j v_j \partial_j B_i - \sum_j B_j
\partial_j v_i$. We then proceed through the same steps as before and replace
the combination of velocity and derivation by a random variable in which we
absorb the dissipative term and the right hand side, remove dimensions, and
model saturation through a cubic term. This leads to the following system:
\begin{equation}
\partial_t B_i +(\alpha_1(t)+\alpha_2(t)+\alpha_3(t)) B_i -
\alpha_i(t) (B_1+B_2+B_3) = -B_i^3 ~~~,~~~ i=1,2,3,
\label{3Dmodel}
\end{equation}
in which the $\alpha_i(t)$ are random variables. This system bears only a
metaphorical relation with the original induction equation and will be used to
exemplify two limiting cases:

If the fluctuations of the $\alpha_i(t)$ are small compared with the mean of the
$\alpha_i(t)$, the solution of (\ref{3Dmodel}) will be close to the solution of
the time independent system in which each $\alpha_i(t)$ in (\ref{3Dmodel}) is
replaced by its mean $\langle\alpha_i\rangle$. Let us assume 
$\langle\alpha_1\rangle = \langle\alpha_2\rangle = \langle\alpha_3\rangle < 0$. An eigenvalue analysis of the left hand
side of (\ref{3Dmodel}) then reveals one neutral mode and two
modes with equal and positive growth rate. In
the presence of small fluctuations, the neutral mode will not contribute significantly
to the dynamics. If the initial conditions and the nonlinear term select an
arbitrary direction in the space spanned by the two degenerate growing modes, we
expect (\ref{3Dmodel}) to behave the same as the single mode model. This is a
fortiori true for a dynamical system with a single non-degenerate growing mode.

If on the other hand the fluctuations of the $\alpha_i(t)$ are large compared
with their means, the dynamics is not dominated by a single mode anymore and the
analysis of the previous section breaks down. The spectrum of the fluctuations
of $\sqrt{\sum B_i^2}$ can now be different and
must be found from numerical computation.

\begin{figure}[h]
\begin{center}
\includegraphics[width=9cm]{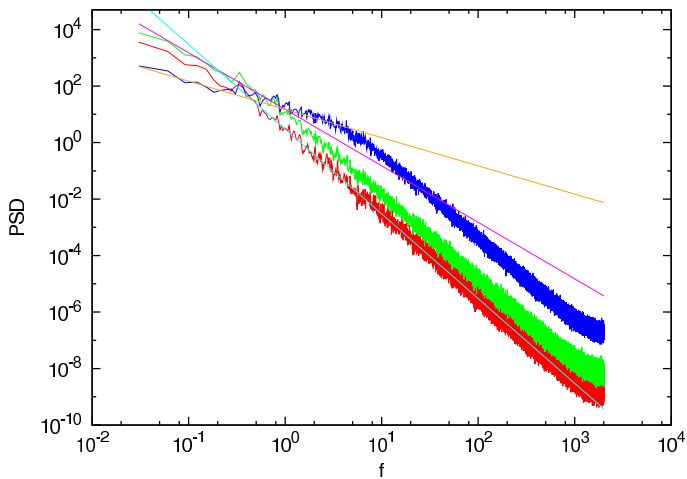}
\end{center}
\caption{Spectral power density of $\sqrt{\sum B_i^2}$ obtained from
the solution of eq. (\ref{3Dmodel}),
as a function of frequency $f=\omega/(2\pi)$, for $\langle\alpha_i^2\rangle=1$ and
$\langle\alpha_i\rangle=-0.01$ (red), -0.3 (green) and -5 (blue).
The straight lines indicate the power
laws $\omega^{-1}$, $\omega^{-2}$ and $\omega^{-3}$.}
\label{fig_3D}
\end{figure}

Fig. \ref{fig_3D} shows some examples of solutions of eq. (\ref{3Dmodel}) in which the
spectrum of the fluctuations of the $\alpha_i(t)$ is in $1/f$ \cite{Kasdin95}
and $\langle\alpha_i^2\rangle=1$. Let us first consider the case in which the
fluctuations of the $\alpha_i$ are small compared with their mean. The spectrum
of $\sqrt{\sum B_i^2}$ shown in fig. \ref{fig_3D} should then behave as
predicted by the single mode model: At the smallest frequencies, the spectrum
of $\sqrt{\sum B_i^2}$ must decay the same as the spectrum of the $\alpha_i$,
i.e. as $1/f$ in the present example. Above a frequency on the order of
$\langle\alpha_i\rangle$, there must be a factor $f^2$ between the power
laws followed by the spectra of $\sqrt{\sum B_i^2}$ and the $\alpha_i$, which
implies a spectrum in $1/f^3$ for $\sqrt{\sum B_i^2}$ in the example considered
here. All these predictions fit well the spectrum shown in fig. \ref{fig_3D} for
$\langle\alpha_i^2\rangle=1$ and $\langle\alpha_i\rangle=-5$.
In order to further
support the applicability of the single mode model, we also computed the angle
between the instantaneous vector $\bi B(t)=(B_1(t),B_2(t),B_3(t))$ and its mean
$\langle\bi B\rangle$. The cosine of that angle,
$\langle\bi B\rangle \cdot \bi B(t) / \sqrt{|\langle\bi B\rangle|^2 ~ |\bi
B(t)|^2}$ in the statistically stationary
state stays larger than 0.99 for $\langle\alpha_i\rangle=-5$ but is scattered 
over a large interval
for $\langle\alpha_i\rangle=-0.01$. In the latter case the fluctuations of the 
$\alpha_i$ are large compared with their mean
and different exponents unrelated to the single mode model become possible. A decay in
$1/f^2$ appears in fig. \ref{fig_3D} which will become important again in connection with
the convectively driven dynamos in spherical shells discussed below.

\subsection{Homogeneous and isotropic turbulence \label{HIT}}
The main focus of this paper is on slow fluctuations, but it is also worthwhile
to have a look at the fast fluctuations in order to see where the differences
are. We also would like to check whether it was reasonable in 
section \ref{single} to assume a spectrum for $\alpha$ which is flat at low frequencies and
steeper at large frequencies. We will specifically look at fluctuations of total
kinetic and magnetic energy. The spectra of these fluctuations must be different
from flow to flow, but we can expect a unique behavior in flows to which the
Kolmogorov phenomenology (K41 in short after \cite{Kolmog41})
of homogeneous and isotropic turbulence applies. We will employ this
phenomenology in a cartesian domain with periodic boundaries, so that it is
useful to expand all fields in Fourier series. Much effort has been spent in the
past to theoretically deduce the wavenumber dependence of the Fourier
coefficients, the best known theory being of course K41. But the Fourier
coefficients also depend on time, or after an additional transformation, on
frequency. Theories such as K41 are concerned with the temporal mean of those
coefficients. The spectrum of temporal fluctuations of those Fourier
coefficients has received much less attention.

This section will present an extension of the Kolmogorov phenomenology which
predicts the spectrum of fluctuations of the total kinetic energy of a flow.
There is no magnetic field involved, but the result is directly applicable to
magnetic field spectra under certain assumptions.

The mean kinetic energy $\langle E_{\rm{kin}}\rangle$ of a turbulent flow is decomposed into
contributions made by wavevectors of modulus $k$ in the form
$\langle E_{\rm{kin}}\rangle=\int_0^\infty \langle E_k\rangle dk$. According to K41, $\langle E_k\rangle \propto k^{-5/3}$.
The number of modes in a thin shell of radius $k$ in spectral space is
proportional to $k^2$, so that the contribution of a single mode with wavevector
$\bi k$ to the total mean kinetic energy is proportional to $k^{-11/3}$. We now
consider the root mean square of the fluctuations of the total kinetic energy: 
$\sigma_{tot}=\langle(E_{\rm{kin}}^2-\langle E_{\rm{kin}}\rangle^2)\rangle^{1/2}$. The angular brackets denote
average over time. The total root mean square is again decomposed into
contributions of different wavevector shells: $\sigma_{tot}=\int_0^\infty
\sigma_k dk$. In order to compute $\sigma_k$, we assume that the fluctuations of
the $k^2$ modes in the wavevector shell of radius $k$ are all uncorrelated so
that the mean squares of the fluctuations of each single mode simply add to give
$\sigma_k^2$.

We next invoke the concept of self-similarity, which is a central tenant of the
K41 theory: All structures (or eddies or modes) in the inertial range are
statistically indistinguishable from each other after a rescaling of length and
time. This implies that the histogram of the fluctuations in every mode has the
same shape if the fluctuations are given as multiples of the mean amplitude. We
then have to conclude that the root mean square of the fluctuations scales the
same as the mean amplitude, i.e. is proportional to $k^{-11/3}$. It follows that
$\sigma_k \propto \sqrt{k^2} k^{-11/3} = k^{-8/3}$.

Note that it is not possible to deduce the scaling of $\sigma_k$ from the usual
dimensional arguments of K41. The ratio $\sigma_k/\langle E_k\rangle$ tends to zero if 
the number of modes in a wavevector shell tends to infinity because the
fluctuations of the modes in that shell average out to zero. The rms of the total 
fluctuation depends therefore on the number of degrees of freedom,
which in turn depends on the ratio of integral to dissipative length scale.
These two length scales are assumed to be irrelevant in the K41 theory, however.

The energy of the modes in the wavevector shell of radius $k$ is now written in
the form
\begin{equation}
E_k(t)=\langle E_k\rangle + \sigma_k h_k(t)
\end{equation}
in which $h_k(t)$ has to obey $\langle h_k\rangle =0$ and $\langle h_k^2\rangle =1$. The Fourier transform
of the fluctuations of the total energy is
\begin{equation}
\int (E_k(t)-\langle E_k\rangle) e^{-i \omega t} dt \propto \int \sigma_k \hat{h}_k(\omega) dk
\end{equation}
with $\hat{h}_k(\omega) \propto \int h_k(t) e^{-i \omega t} dt$.
We now use again the hypothesis of self-similarity and note that the typical
time scale for a mode with wavenumber $\bi k$ is its turn-over time which is
proportional to $k^{-2/3}$ \cite{Tennek72, Pope00}. Restricting ourselves to
frequencies $\omega$ and wavenumbers $k$ in the inertial range, we expect
\begin{equation}
\hat{h}_k(\omega) \propto k^{-1/3} g(\omega k^{-2/3})
\end{equation}
with an unknown but universal function $g$. The form of the argument reflects
that all $\hat{h}_k(\omega)$ should have the same form once $\omega$ is expressed
in multiples of the turn-over frequency, and the amplitude factor $k^{-1/3}$
follows from the requirement that $\langle h_k^2\rangle=1$, which implies that 
$\int |\hat{h}_k(\omega)|^2 d\omega$ is a constant independent of $k$.
Substituting $z=\omega k^{-2/3}$ into
$\int |\hat{h}_k(\omega)|^2 d\omega = \int k^{-2/3} |g(\omega k^{-2/3}|^2)
d\omega$
shows that this is indeed fulfilled.

We can now proceed to finally compute the spectrum of the energy fluctuations,
using the same substitution $z=\omega k^{-2/3}$, to find:
\begin{equation}
\int (E_k(t)-\langle E_k\rangle) e^{-i \omega t} dt \propto 
\int k^{-8/3} k^{-1/3} g(\omega k^{-2/3}) dk \propto
\omega^{-3} \int z^2 g(z) dz.
\end{equation}
The spectrum of the energy fluctuations, i.e. the square of the Fourier spectrum
above, must thus decay as $\omega^{-6}$ for $\omega$ in the inertial range. This
prediction will be verified below.

Spectra in MHD turbulence can have a variety of shapes, depending on the
presence of Alfv\'en waves, an applied magnetic field, the magnetic Prandtl
number etc. \cite{Verma04, Goldre09} The above calculation can be directly reproduced for the magnetic
energy as long as the magnetic energy density follows a Kolmogorov spectrum in
$k^{-5/3}$. This happens in homogenous and isotropic flows at high magnetic
Reynolds numbers without applied external field \cite{Biskam00}. In that case,
magnetic field fluctuations should decay as $\omega^{-6}$, too.

\section{The G.O. Roberts dynamo \label{Roberts}}
Dynamos based on periodic flows first investigated by G.O. Roberts
\cite{Robert72} have been useful for a number of basic studies of the dynamo
effect \cite{Tilgne04, Tilgne07d, Gubbin07, Tilgne08, Tilgne08b}
and have inspired the Karlsruhe dynamo experiment
\cite{Stiegl01}. In order to verify the ideas developed in the previous section,
we wish to numerically simulate the Navier-Stokes and induction equations in
the following non-dimensional form:
\begin{eqnarray}
\partial_t\bi B + \nabla \times (\bi B \times \bi u) = \frac{\mathrm{Pm}}{\mathrm{Re}} \nabla^2 \bi B \\
\label{eq_induc}
\nabla \cdot \bi B = 0 \\
\partial_t\bi u + \bi u \cdot \nabla \bi u = -\nabla p +\frac{1}{\mathrm{Re}} \nabla^2
\bi u + \bi F + (\nabla \times \bi B) \times \bi B \\
\label{NS}
\nabla \cdot \bi u = 0 
\end{eqnarray}

$\mathrm{Re}$ and $\mathrm{Pm}$ are two control parameters standing for
Reynolds and magnetic Prandtl numbers, respectively. The forcing $\bi F$
will take the form
\begin{equation}
\bi F = \frac{8 \pi^2}{\mathrm{Re}} \bi u_0
\label{F}
\end{equation}
One solution to the above equations is then $\bi B=0$, $\bi u = \bi u_0$ with
\begin{eqnarray}
\bi u_0=
\left( \begin{array}{ll}
\sqrt{2} \sin (2 \pi x) \cos (2 \pi y)\\
-\sqrt{2} \cos (2 \pi x) \sin (2 \pi y)\\
2 \sin (2 \pi x) \sin (2 \pi y)\\
\end{array} \right)
\label{eq_flowI}
\end{eqnarray}
This solution is unstable for large enough $\mathrm{Re}$, either because the
flow is hydrodynamically unstable or because the dynamo effect leads to $\bi B
\neq 0$. The velocity
field (\ref{eq_flowI}) is a cartesian arrangement of helical eddies with their
axes along $z$. Periodic boundary conditions will be applied in all cartesian
directions $x$, $y$, $z$ with periodicity lengths $l_x=1$, $l_y=1$, and three
different choices of $l_z$, namely $l_z=1$, $1.5$ and 2.

If we call in dimensional units $L_x$ the periodicity length in $x-$direction,
the laminar velocity $\bi v_0 = V_0 \bi u_0$, and the kinematic viscosity and
magnetic diffusivity $\nu$ and $\lambda$, respectively, the nondimensional
parameters are given by $\mathrm{Pm}=\nu/\lambda$ and $\mathrm{Re}=V_0 L_x/\nu$.

The above problem is conveniently discretized with a spectral method. However,
we need very long time series if we are interested in small frequencies and the
availability of computer hardware for long runs becomes an important
consideration in the choice of the numerical method. The most readily available
platform happened to be a set of machines equipped with CUDA (compute unified
device architecture) capable graphic
processing units. These architectures are not well adapted to spectral methods,
nor to Poisson solvers. That is why a fully explicit finite difference scheme
was implemented to solve the following set of equation:
\begin{eqnarray}
\partial_t\bi B + \nabla \times (\bi B \times \bi u) = \frac{\mathrm{Pm}}{\mathrm{Re}}
\nabla^2 \bi B \\
\label{eq_induc2}
\nabla \cdot \bi B = 0 \\
\partial_t\bi u + \bi u \cdot \nabla \bi u = -c^2 \nabla \rho +\frac{1}{\mathrm{Re}}
\nabla^2
\bi u + \bi F + (\nabla \times \bi B) \times \bi B \\
\label{NS2}
\partial_t \rho + \nabla \cdot \bi u = 0 
\end{eqnarray}
These equations describe a compressible fluid with the equation of state 
$p = c^2 \rho$ and a linearized continuity equation. Solutions of the
incompressible Navier-Stokes equation will be recovered in the limit of the
sound speed $c$ tending to infinity. The sound speed was chosen large enough so
that the peak Mach number never exceeded $0.12$ in the computations presented
below. The typical Mach number was more like $0.05$.
The finite difference code was validated against a spectral code
\cite{Sarkar06} by comparing the results of short runs, but all the heavy
computation was done with the finite difference code.

\begin{figure}[h]
\begin{center}
\includegraphics[width=9cm]{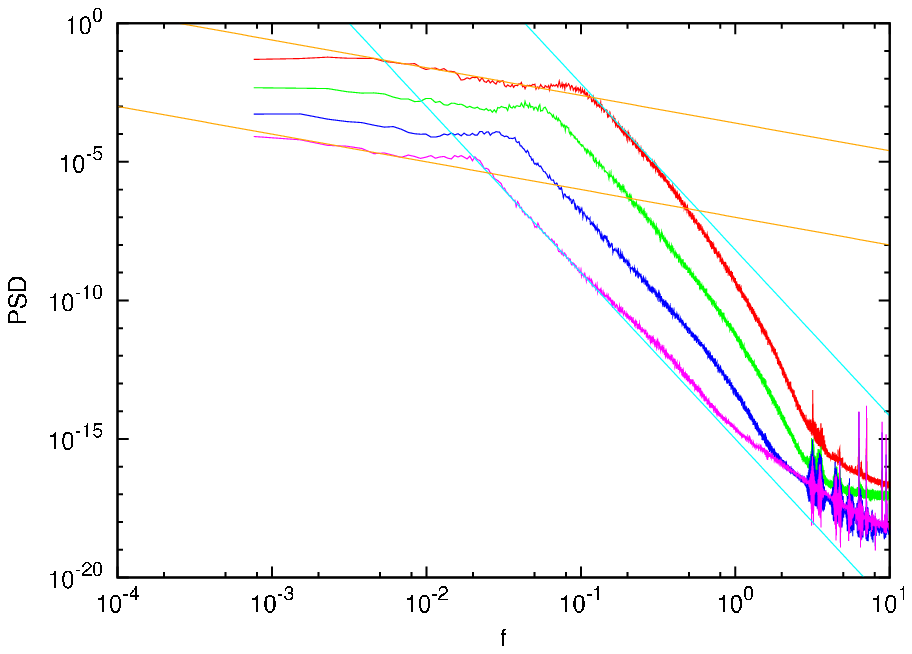}
\includegraphics[width=9cm]{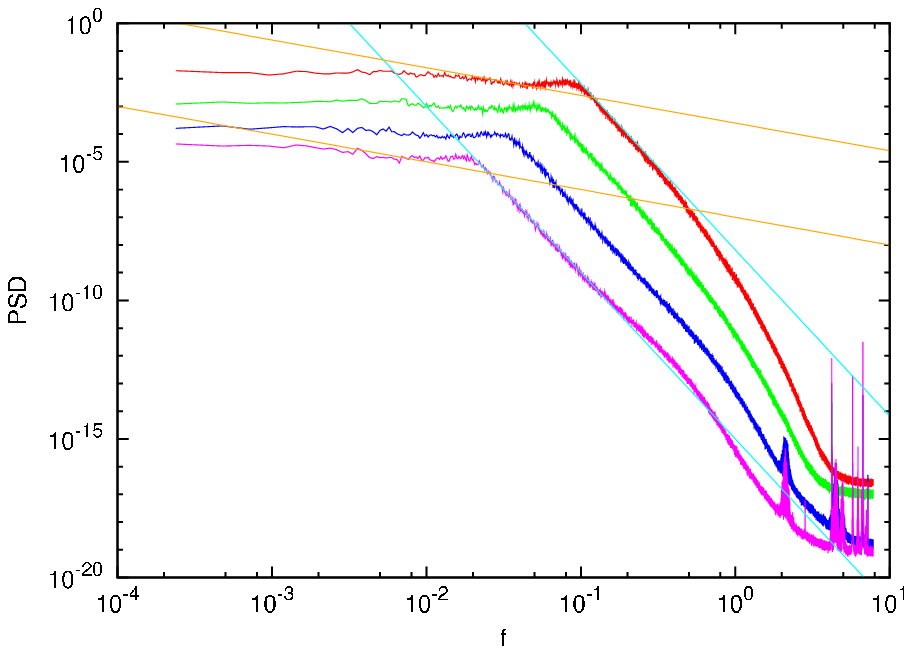}
\includegraphics[width=9cm]{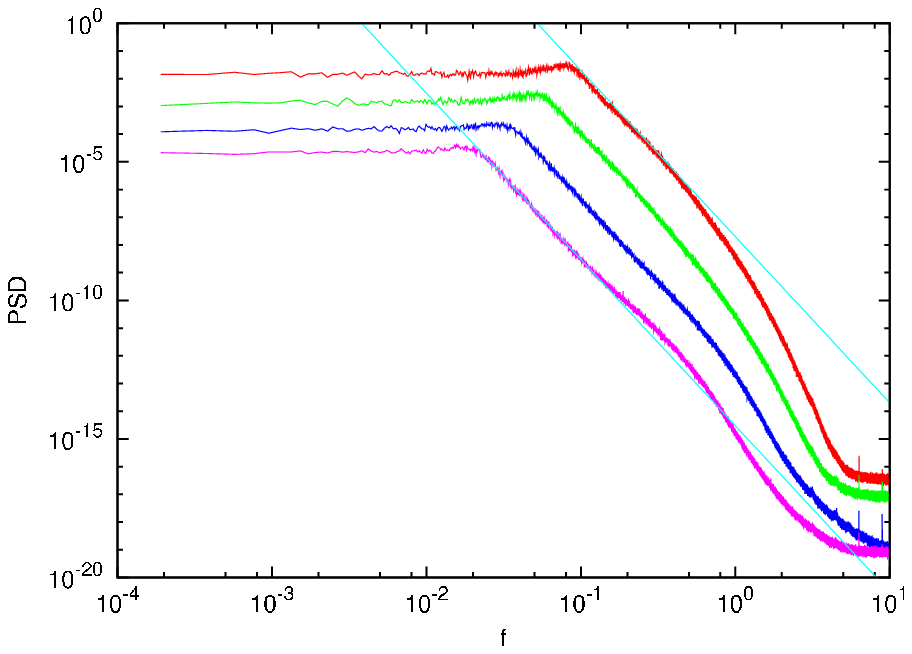}
\end{center}
\caption{Spectral power density of the kinetic energy as a function of frequency
$f=\omega/(2\pi)$ for flows without magnetic field. The panels show spectra for 
$l_z=2$, $1.5$ and 1 and each panel contains spectra for $\mathrm{Re}=10^3$ (red), $3
\times 10^3$ (green), $10^4$ (blue) and $3 \times 10^4$ (pink). The straight lines indicate the power
laws $\omega^{-1}$ and $\omega^{-6}$.}
\label{fig_Ekin}
\end{figure}

We will first look at non-magnetic flows. Figure \ref{fig_Ekin} shows spectra of
kinetic energy density $E_{\rm{kin}}$ for $l_z=2$, $1.5$ and 1. $E_{\rm{kin}}$ is defined
as 
\begin{equation}
E_{\rm{kin}}=\frac{1}{2V}\int \bi u^2 dV,
\end{equation}
where $V$ is the computational volume.
In all cases, the spectra of $E_{\rm{kin}}$ follow an $\omega^{-6}$ over at least a
decade in $\omega$. This power law identifies the inertial range of a Kolmogorov
cascade according to section \ref{HIT} (It was not checked directly whether the
spatial spectrum of the velocity fields follows the K41 law because the
simulations were not done with a spectral code and the evaluation of the spatial
spectra would have been cumbersome). An even steeper decay follows at higher
frequencies before the spectrum dives under the noise level introduced by roundoff
error. 

On the low frequency side, the inertial range ends in a conspicuous hump.
The frequency of the local maximum in the spectrum, $f_{\rm{max}}$, scales with what
one may identify as the energy injection scale: $\mathrm{Re}$ is a control
parameter in eq. (\ref{NS}), but the Reynolds number determined a posteriori is
$\mathrm{Re}_2=\sqrt{E_{\rm{kin}}}$. The values of $\mathrm{Re}_2$ given in table
\ref{table1} vary from 340 to 2100 indicating that all flows are turbulent. The
product $\mathrm{Re}_2 f_{\rm{max}}$ varies from 27 to 41. Despite this variation,
we identify $f_{\rm{max}}$ with the injection scale because the definition of
$\mathrm{Re}_2$ does not take into account that the flow is anisotropic so that
we do not expect $\mathrm{Re}_2 f_{\rm{max}}$ to be strictly constant.

\begin{table}\centering
\begin{tabular}{|c|c|c|c|}

\hline
$l_z$ & $\mathrm{Re}$ & $\mathrm{Re}_2$ & $\mathrm{Re}_2f_{\rm{max}}$  \\\hline
\hline

2 & $10^3$ & 341 & 27 \\
2 & $3 \times 10^3$ & 630 & 29 \\
2 & $10^4$ & 1171 & 33 \\
2 & $3 \times 10^4$ & 2038 & 41 \\
\hline

1.5 & $10^3$ & 353 & 28 \\
1.5 & $3 \times 10^3$ & 640 & 33 \\
1.5 & $10^4$ & 1191 & 32 \\
1.5 & $3 \times 10^4$ & 2078 & 37 \\
\hline

1 & $10^3$ & 382 & 31 \\
1 & $3 \times 10^3$ & 670 & 35 \\
1 & $10^4$ & 1235 & 35 \\
1 & $3 \times 10^4$ & 2160 & 37 \\
\hline

\end{tabular}
\caption{$l_z$, $\mathrm{Re}$, $\mathrm{Re}_2$ and $\mathrm{Re}_2f_{\rm{max}}$ for the 
simulations in fig. \ref{fig_Ekin}.}
\label{table1}
\end{table}

\begin{figure}[h]
\begin{center}
\includegraphics[width=9cm]{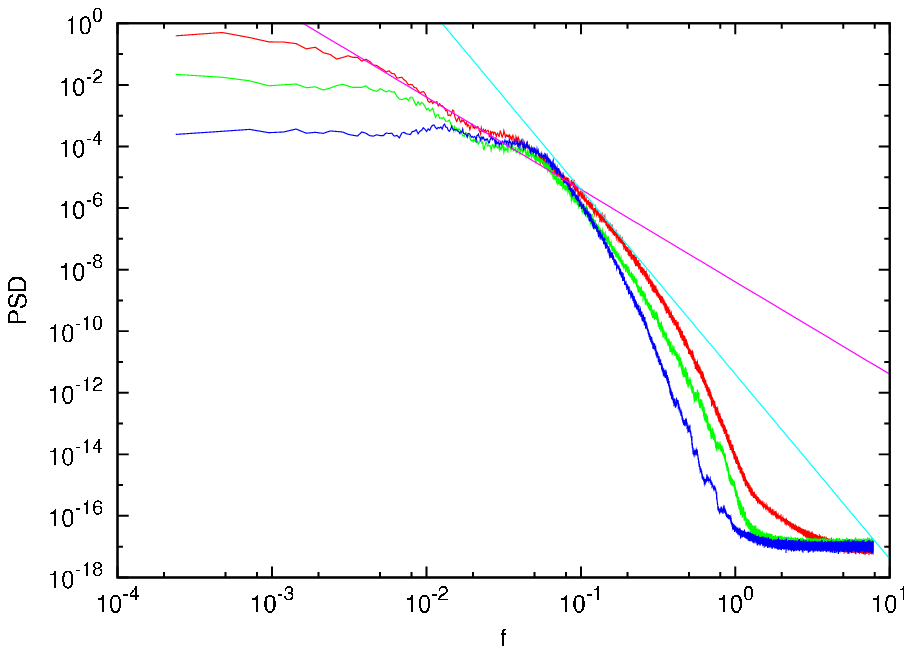}
\includegraphics[width=9cm]{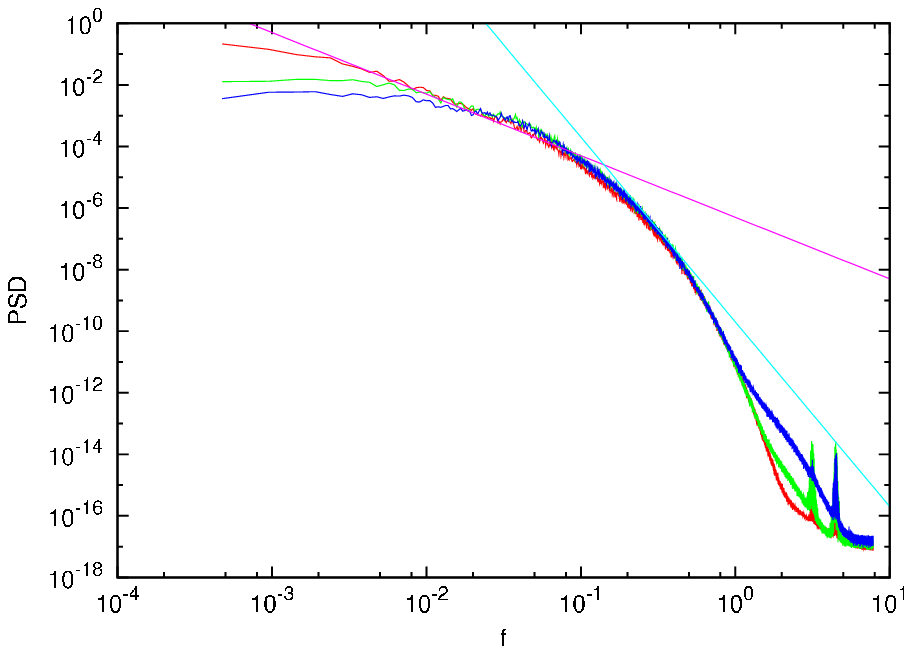}
\end{center}
\caption{Spectral power density of the magnetic energy $E_B$ as a function of frequency
$f=\omega/(2\pi)$. The panels show spectra for $l_z=1.5$ (top) and 1 (bottom) and
$\mathrm{Re}=10^4$. The different lines in the top are for $\mathrm{Pm}=0.03$ (red), 
0.05 (green) and 0.1 (blue), and in the bottom panel for $\mathrm{Pm}=0.27$ (red), 0.5 (green)
and 1 (blue). The straight lines indicate the power laws $\omega^{-3}$ and $\omega^{-6}$
(top) and $\omega^{-2}$ and $\omega^{-6}$ (bottom).}
\label{fig_EB}
\end{figure}

The spectra at frequencies smaller than the injection frequency are different
for $l_z=1$ and $l_z=2$ or $1.5$: For $l_z=1$, the spectrum is purely white
noise, whereas a section of $1/f$-noise is visible for $l_z=2$ and $1.5$. The
section of $1/f$-noise shrinks with increasing $\mathrm{Re}$.

We next turn to dynamos. For $\mathrm{Pm}$ slightly above the critical value
given in table \ref{table1}, the spectra of the magnetic energy density $E_B$,
defined as $E_B=\frac{1}{2V} \int \bi B^2 dV$, show a segment of spectrum in
$\omega^{-3}$ for $l_z=2$ and $1.5$ and in $\omega^{-2}$ for $l_z=1$. A white
spectrum is always found at the smallest frequencies (figure \ref{fig_EB}).
This matches the prediction of section \ref{single} in as far as the velocity
spectrum without magnetic field contains an interval with a spectrum close to
$\omega^{-1}$ for $l_z=2$ and $1.5$ and nothing but white noise below the
injection frequency for $l_z=1$. According to section \ref{single}, there should
be a factor $\omega^2$ between magnetic and velocity spectra at low frequencies,
which is compatible with the results in figures \ref{fig_Ekin} and \ref{fig_EB}.

The transition from white noise to either $\omega^{-2}$ or $\omega^{-3}$ occurs
at a higher frequency for larger $\mathrm{Pm}$ in figure \ref{fig_EB}.
This is again in agreement with
section \ref{single} because a larger $\mathrm{Pm}$ at constant $\mathrm{Re}$
corresponds to a flow
farther above onset, which corresponds to a larger $\langle\alpha\rangle$ in the model of
section \ref{single}.

\begin{figure}[h]
\begin{center}
\includegraphics[width=9cm]{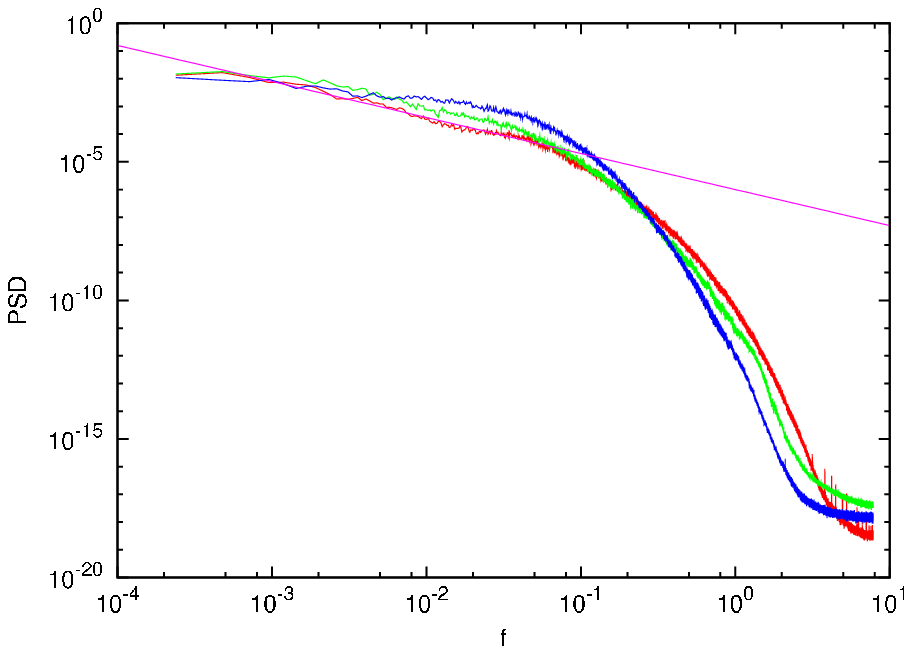}
\includegraphics[width=9cm]{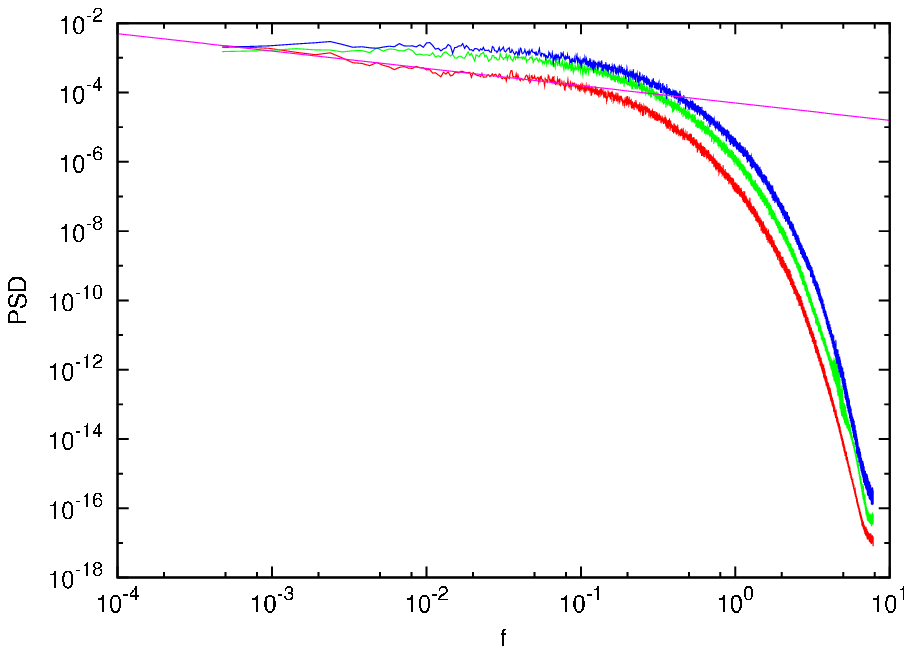}
\end{center}
\caption{Spectral power density of the local magnetic field amplitude $|\bi B|$
at the position $x=y=0$, $z=l_z/2$ as a function of frequency
$f=\omega/(2\pi)$. The panels show spectra for $l_z=1.5$ (top) and 1 (bottom) with
$\mathrm{Re}=10^4$ and the same $\mathrm{Pm}$ as in fig. \ref{fig_EB}. 
The straight lines indicate the power laws $\omega^{-1.3}$ (top) and $\omega^{-1/2}$ (bottom).}
\label{fig_Bloc}
\end{figure}

Experiments use local probes to characterize the magnetic field. In the present
simulations, the magnitude $|\bi B|$ of the field at position $x=y=0$, $z=l_z/2$
was recorded as a function of time. This location corresponds to a position at
which the magnetic field has been measured in the Karlsruhe experiment
\cite{Muller04}. The spectra of the magnetic field amplitude in one point show a
power law at low frequencies over a larger interval than the spectra of $E_B$,
but with an exponent which bears no obvious relation with the exponent of the
energy spectra (figure \ref{fig_Bloc}). There does not seem to be any
theoretical tool for predicting local fluctuations, so that we simply note the
spectra in figure \ref{fig_Bloc} as an empirical fact. However, they also
approach a white noise spectrum as $\mathrm{Pm}$ increases, just as the magnetic
energy spectra do in fig. \ref{fig_EB}.

\section{Dynamos in spherical shells \label{geo}}
Simulations of convectively driven dynamos in spherical shells reported in
\cite{Olson07} yielded spectra with power laws which are not compatible with the
single mode model. According to the phenomenology developed in section
\ref{several}, this simply means that the fluctuations in these dynamos are
large. It is expected that at sufficiently low Rayleigh numbers and in
sufficiently quiet convection, one finds again spectra in agreement with the
single mode model. In order to test this idea, we also performed simulations of
dynamos in spherical shells. Because these computations are more expensive, the
spectra are more noisy and not as extended in frequency as for the G.O. Roberts
dynamos, but they demonstrate the main effect.

We consider the same physical model as in \cite{Olson07} of a rotating spherical
shell with its gap filled with fluid driven by convection. The variable
quantifying buoyancy will be called temperature here, but the employed boundary
conditions better fit compositional convection: Fixed temperature at the inner
boundary and zero heat flux through the outer boundary. Both boundaries are
assumed no slip and electrically insulating.
The following dimensionless equations 
\begin{eqnarray}
\partial_t \bi{u}+(\bi{u} \cdot
{\nabla})\bi{u}+2\hat{z}\times\bi{u} = \nabla\phi-\frac{\mathrm E^2
\mathrm{Ra}}{\mathrm{Pr}}T\frac{\bi{r}}{r_0}+(\nabla\times\bi{B})\times\bi{B}+\mathrm E\nabla^2\bi{u}
 \\
\partial_t T+(\bi{u} \cdot {\nabla}) T = \frac{\mathrm E}{\mathrm{Pr}}\nabla^2
T- \mathrm E \epsilon 
 \\
\partial_t \bi{B}+\nabla\times(\bi{B}\times\bi{u}) =
\frac{\mathrm E}{\mathrm{Pm}}\nabla^2\bi{B} 
\\
{\nabla}\cdot \bi{u}=0\mbox{      } \mbox{  ,    }  \mbox{      }
{\nabla}\cdot \bi{B}=0 
\end{eqnarray}
are solved in a spherical shell
for the velocity, magnetic and temperature fields $\bi u$, $\bi B$
and $T$. The control parameters are the Ekman number $\mathrm E$, the Rayleigh number
$\mathrm{Ra}$, the Prandtl number $\mathrm{Pr}$, and the magnetic Prandtl number
$\mathrm{Pm}$. The ratio of inner and outer radii is fixed at $0.35$ with the
dimensionless outer radius being $r_0=1/0.65$, and
$\epsilon$, the variable modeling the buoyancy source \cite{Olson07} is fixed at
$\epsilon =1$. The equations have been solved using the spectral method described
in \cite{Tilgne99d} with a resolution of 33 Chebychev polynomials in radius and
spherical harmonics of degree up to 64.

\begin{figure}[h]
\begin{center}
\includegraphics[width=9cm]{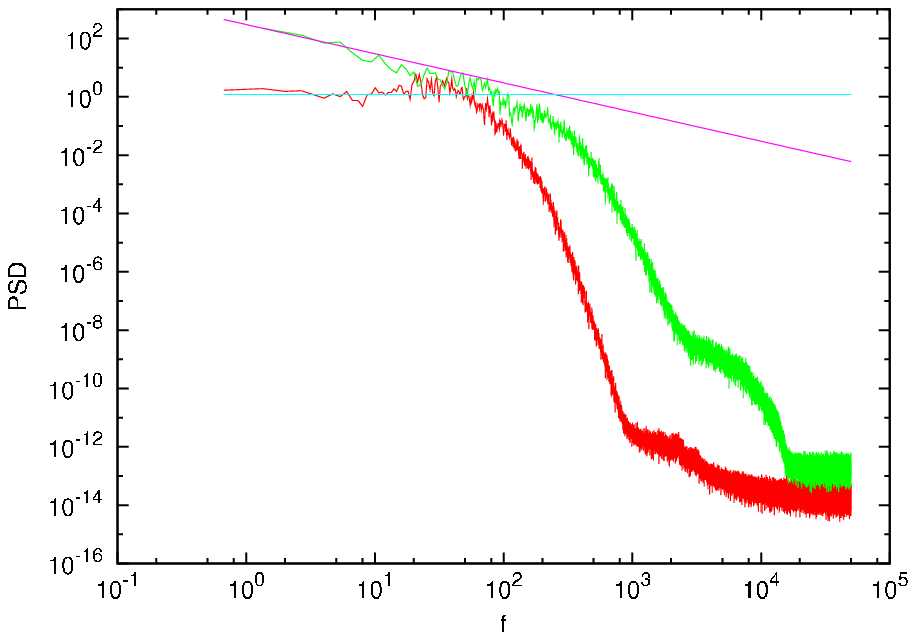}
\includegraphics[width=9cm]{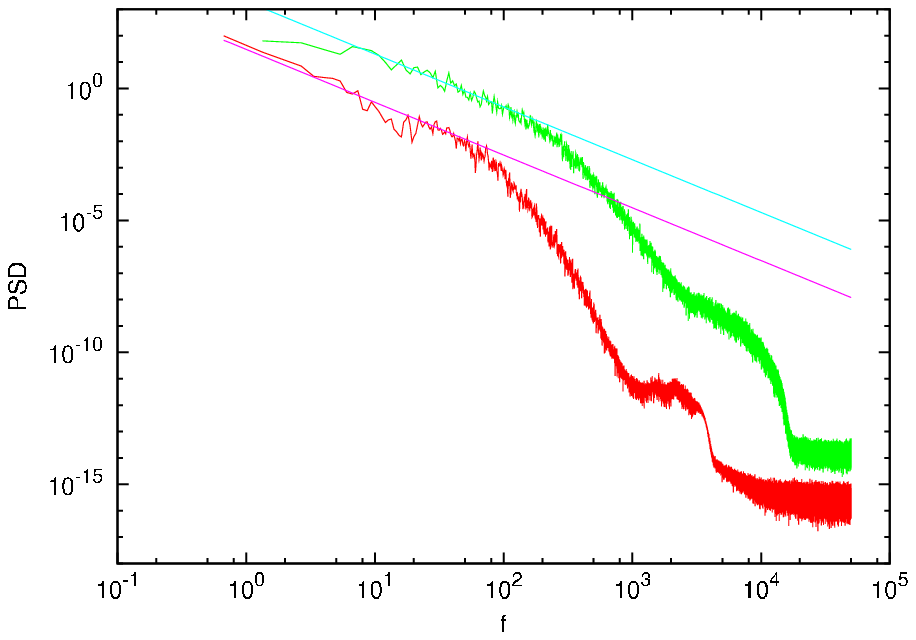}
\end{center}
\caption{Spectral power density of the kinetic energy $E_{\rm{kin}}$ (top) and the magnetic
energy $E_B$ (bottom) as a function of frequency $f=\omega/(2\pi)$
for $\mathrm{Pm}=13$, $\mathrm{Ek}=6.5 \times 10^{-3}$, $\mathrm{Pr}=1$ and
$\mathrm{Ra}=10^5$ (red) or $\mathrm{Ra}=10^6$ (green). The straight lines indicate the
power laws $f^0$ and $f^{-1}$ in the top panel and $f^{-2}$ in the bottom panel.}
\label{fig_sphere}
\end{figure}

Fig. \ref{fig_sphere} shows two calculations in which all parameters are held
constant except for the Rayleigh number. In going from the low to the high
Rayleigh number, the spectrum of kinetic energy changes from white noise to
approximately $f^{-1}$. The spectrum of magnetic energy shows a decay in
$f^{-2}$ in both cases. This means that at low $\mathrm{Ra}$, the exponents in
the power laws for the kinetic and magnetic energies differ by 2, just as they
should according to the single mode model. At higher $\mathrm{Ra}$, i.e. in the
flow with larger fluctuations, the exponents are different as expected from
section \ref{several} and differ by 1. Incidentally, the
combination of $f^{-1}$ and $f^{-2}$ for kinetic and magnetic energies is identical
with a combination of exponents found in the low dimensional dynamic system of
section \ref{several}.

\section{Conclusion}

Colored noise has been noted in fluctuations of magnetic field strength in both
experiments and simulations. According to common experience with turbulent
velocity fields, the fluctuations of mechanical quantities have on the contrary
a white noise at low frequencies, a low frequency being a frequency smaller
than the inverse of the integral time scale. It has been shown in the present
paper that even the simplest phenomenology based on a single mode model predicts
a spectrum in $1/f^2$ for the total magnetic energy if the velocity field is
characterized by white noise. The single mode model is justified as long as the
fluctuations in the velocity field are small enough in amplitude. These simple
models are corroborated by numerical solutions of the dynamo equations for 2D
periodic dynamos and for convectively driven dynamos in spherical shells.

Colored noise must have a low frequency cut-off for the power integral to
converge. This low frequency is independent of diffusive time scales according
to the models studied here. Instead, it depends on how strongly a dynamo is
driven. The more supercritical a dynamo is, the larger is the cut-off frequency.

It is also possible to deduce the spectrum of turbulent magnetic fluctuations
from the Kolmogorov phenomenology. Within the inertial range, fluctuations of
the total kinetic and magnetic energy should decay as $f^{-6}$, which is verified by
numerical simulation.

It is left to future work to deal with fluctuations of the magnetic field at a
given point in space. If the magnetic field is dominated by a large scale component,
local fluctuations are of course similar to global fluctuations. At large
magnetic Reynolds numbers, small scale fluctuations exceed in magnitude the
fluctuations of the large scale field and the spectrum of the fluctuations of local
magnetic fields can be different from the spectrum of total magnetic energy. The
spectra computed for the 2D periodic dynamos at high magnetic Reynolds numbers
reveal power laws at low frequencies in the range $f^{-1.3}$ to $f^{-0.5}$.

\section*{References}
%\bibliographystyle{unsrt}
%\bibliography{dynamo}

\end{document}